\documentstyle[aps,prl,preprint,floats,epsfig]{revtex}

\begin{document}

\preprint{\tighten\vbox{\hbox{\hfil CLNS 98/1541}
                        \hbox{\hfil CLEO 98-2}
}}

\title{First Observation of the Cabibbo Suppressed 
Decay $B^+\rightarrow \bar D^0 K^+$}

\author{CLEO Collaboration}
\date{\today}

\maketitle
\tighten

\begin{abstract}
We have observed the decay $B^+\rightarrow \bar D^0 K^+$, using
3.3~million $B\bar B$ pairs collected with the CLEO~II detector
at the Cornell Electron Storage Ring. We find the ratio of branching
fractions $R\equiv{\cal B}(B^+\rightarrow \bar D^0 K^+) / {\cal
B}(B^+\rightarrow \bar D^0 \pi^+) = 0.055 \pm 0.014 \pm 0.005$.
\end{abstract}
\newpage

{
\renewcommand{\thefootnote}{\fnsymbol{footnote}}



\begin{center}
M.~Athanas,$^{1}$ P.~Avery,$^{1}$ C.~D.~Jones,$^{1}$
M.~Lohner,$^{1}$ S.~Patton,$^{1}$ C.~Prescott,$^{1}$
J.~Yelton,$^{1}$ J.~Zheng,$^{1}$
G.~Brandenburg,$^{2}$ R.~A.~Briere,$^{2}$
% ...
\end{center}
 
\small
\begin{center}
$^{1}${University of Florida, Gainesville, Florida 32611}\\
$^{2}${Harvard University, Cambridge, Massachusetts 02138}\\
% ...
\end{center}
 
\setcounter{footnote}{0}
}
\newpage

\begin{figure}[p]
\centering
\leavevmode
\epsfxsize=5.00in
\epsffile{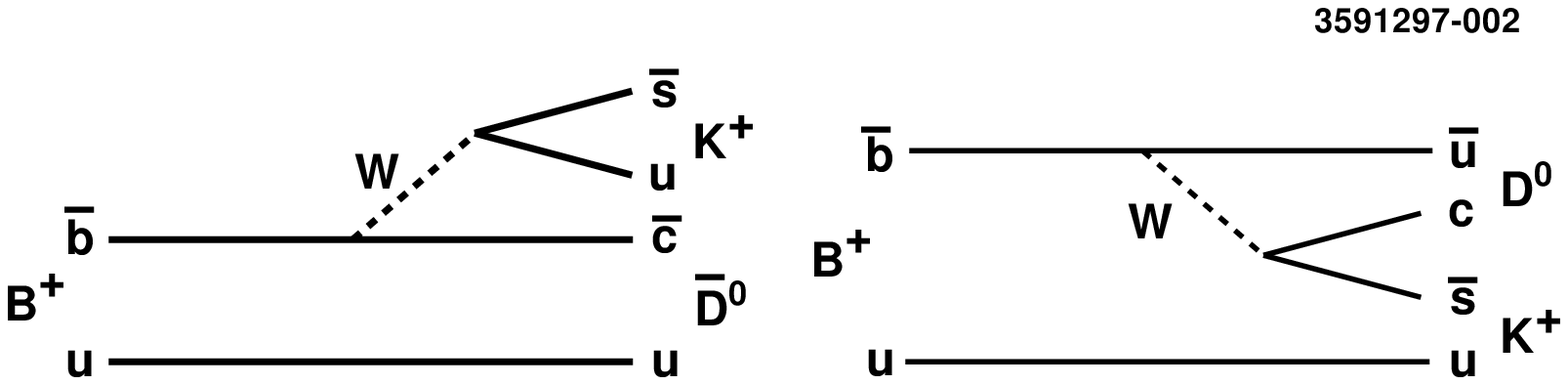}
\caption{Dominant Feynman diagrams of $B^+\rightarrow \bar D^0 K^+$ and
$B^+\rightarrow D^0 K^+$.}
\label{fig:b2dk}
\end{figure}

Several authors~\cite{ref:b2dk} have devised methods for measuring the
phase $\gamma \approx \arg(V^*_{ub})$ of the Cabibbo-Kobayashi-Maskawa
(CKM)~\cite{ref:ckm} unitarity triangle, using decays of the type
$B\rightarrow DK$.  Comparison between these measurements and results
from other $B$ and $K$ decays may be used to test the CKM model of
$CP$~violation.  $CP$~violation could be manifested in $B\rightarrow DK$ in
the interference between a $\bar b\rightarrow \bar c$ and a
$\bar b\rightarrow \bar u$ amplitude
(Figure~\ref{fig:b2dk}), detected when the $D$~meson is observed in a
final state accessible to both $D^0$ and $\bar D^0$.


The data used in this analysis were produced in $e^+e^-$ annihilations
at the Cornell Electron Storage Ring (CESR), and collected with the
CLEO~II detector~\cite{ref:detector}. The data consist of $3.1~{\rm
fb}^{-1}$ taken at the $\Upsilon$(4S) resonance, containing approximately
3.3~million $B\bar B$ pairs. To study the continuum $e^+e^-
\rightarrow q\bar q$ background, we use $1.6~{\rm fb}^{-1}$ of
off-resonance data, taken 60~MeV below the $\Upsilon$(4S) peak.

CLEO~II is a general-purpose solenoidal magnet detector.
The momenta of charged particles are measured in a tracking system,
consisting of a 6-layer straw tube chamber, a 10-layer precision drift
chamber, and a 51-layer main drift chamber, all operating inside a 1.5
T superconducting solenoid.  The main drift chamber also provides
measurements of the specific ionization, $dE/dx$, which we use for
particle identification. Photons are detected in a 7800-CsI crystal
electromagnetic calorimeter inside the magnet coil. Muons are
identified using proportional counters placed at various depths in the
magnet return iron.

We reconstruct $\bar D^0$ candidates in the decay modes $K^+\pi^-$,
$K^+\pi^-\pi^0$, or $K^+\pi^-\pi^+\pi^-$ (reference to the
charge-conjugate state is implied). Pion and kaon candidate tracks are
required to originate from the interaction point and satisfy criteria
designed to reject spurious tracks. Muons are rejected by requiring
that the tracks stop in the first five interaction lengths of the muon
chambers. Electrons are rejected using $dE/dx$ and the ratio of the
track momentum to the associated calorimeter shower energy.  The $\bar
D^0$ daughter tracks are required to have $dE/dx$ consistent with
their particle hypothesis to within three standard
deviations~($\sigma$). Neutral pion candidates are reconstructed from pairs
of isolated calorimeter showers with invariant mass within 15~MeV
(approximately $2.5\sigma$) of the nominal $\pi^0$ mass. The lateral
shapes of the showers are required to be consistent with those of
photons. We require a minimum energy of 30~MeV for showers in the
barrel part of the calorimeter, and 50~MeV for endcap showers. At
least one of the two $\pi^0$ showers is required to be in the
barrel. The $\pi^0$ candidates are kinematically fitted with the
invariant mass constrained to be the $\pi^0$ mass.

The invariant mass of the $\bar D^0$ candidate,
$M(D)$, is required to be within 60~MeV of the nominal $\bar D^0$
mass. The $M(D)$~resolution, $\sigma_{M(D)}$, is 9~MeV in the
$K^+\pi^-$ mode, 13~MeV in the $K^+\pi^-\pi^0$ mode, and 7~MeV in the
$K^+\pi^-\pi^+\pi^-$ mode.  The loose $M(D)$~requirement leaves a
broad sideband to assess the background.

$B^+$~candidates are formed by combining a $\bar D^0$~candidate with
a ``hard'' kaon candidate track. For each $B^+$ candidate, we calculate the
beam-constrained mass, $M_{bc} \equiv \sqrt{E_{\rm b}^2 - p_B^2}$, where
$p_B$ is the $B^+$ candidate momentum and $E_{\rm b}$ is the beam
energy. $M_{bc}$ peaks at the nominal $B^+$ mass for signal, with a
resolution of $\sigma_{M_{bc}} = 2.6$~MeV, determined mostly by the beam
energy spread. We accept candidates with $M_{bc} > 5.230$~GeV.
We define the energy difference,
$\Delta E \equiv E_D + \sqrt{p_K^2 + M_K^2}- E_{\rm b}$, where $E_D$
is the measured energy of the $\bar D^0$ candidate, $p_K$ is the
momentum of the hard kaon candidate, and $M_K$ is the nominal kaon
mass. Signal events peak around $\Delta E = 0$, with a resolution of
24~MeV in the $K^+\pi^-$ mode, 27~MeV in the $K^+\pi^-\pi^0$ mode, and
20~MeV in the $K^+\pi^-\pi^+\pi^-$ mode.  We require $-100 < \Delta E
< 200$~MeV.


The largest source of background is the Cabibbo allowed decay
$B^+\rightarrow \bar D^0 \pi^+$, distributed around $\Delta E =
48$~MeV. Taking into account correlations between $\Delta E$ and
$M(D)$, the $\Delta E$ separation between signal and $B^+\rightarrow
\bar D^0 \pi^+$ is about $2.3\sigma$ in all three modes.  The only
additional variable which provides significant $K-\pi$ separation is
$dE/dx$ of the hard kaon candidate. The $dE/dx$ separation between
kaons and pions in the relevant momentum range of $2.1-2.5$~GeV is
approximately $1.5\sigma$. Our $dE/dx$ variable is chosen such that
pions are distributed approximately as a zero-centered, unit-r.m.s.
Gaussian, and kaons are centered around $-1.4$, with a width of about
0.9.

Other sources of $B\bar B$ background are $B\rightarrow \bar D^*
\pi^+$, $B^+\rightarrow \bar D^0 \rho^+$, and events with a
misreconstructed $\bar D^0$ which pass the selection criteria. Such
$B\bar B$ events tend to have low $\Delta E$ and broad $M_{bc}$
distributions.  Continuum $e^+e^- \rightarrow q\bar q$ events also
contribute to the background. We reject 69\% of the continuum and
retain 87\% of the signal by requiring $|\cos\theta_s| < 0.9$, where
$\theta_s$ is the angle between the sphericity axis of the $B^+$
candidate and that of the rest of the event.  The sphericity axis,
${\boldmath \bf s}$, of a set of momentum vectors, $\{ {\boldmath \bf
p}_i \}$, is the axis for which $\sum_i | {\boldmath \bf p}_i \times
{\boldmath \bf s}|^2$ is minimized.

In addition to the above variables, discrimination between signal and
continuum background is obtained from $\cos\theta_B$, where $\theta_B$
is the angle between the $B^+$ candidate momentum and the beam axis,
and by using a Fisher discriminant~\cite{ref:bigrare}. The Fisher
discriminant is a linear combination, ${\cal F}\equiv
\sum_{i=1}^{11}\alpha_i y_i$, where the coefficients $\alpha_i$ are
chosen so as to maximize the separation between $B\bar B$ and
continuum Monte Carlo samples. The eleven variables, $y_i$, are
$|\cos\theta_{thr}|$ (the cosine of the angle between the $B^+$
candidate thrust axis and the beam axis), the ratio of the Fox-Wolfram
moments $H_2/H_0$~\cite{ref:fox}, and nine variables measuring the
scalar sum of the momenta of tracks and showers from the rest of the
event in nine, $10^\circ$ angular bins centered about the candidate's
thrust axis. Signal events peak around ${\cal F}= 0.4$, while
continuum events peak at ${\cal F}= 2$, both with approximately unit
r.m.s.

18.8\% of the events have more than one $B^+$~candidate, reconstructed
in any of the three modes, which satisfies the selection criteria. In
such events we select the best candidate, defined to have the smallest
$\chi^2 \equiv [(M_{bc} - M_B)/\sigma_{M_{bc}}]^2 + [(M(D) -
M_D)/\sigma_{M(D)}]^2$, where $M_B$ and $M_D$ are the nominal $B$~and
$D$~masses, respectively.  We verify that the distribution of the
number of candidates per event in the Monte Carlo agrees well with the
data.

The efficiency of signal events to pass all the requirements is $0.4412 \pm
0.0029$ for the $K^+\pi^-$ mode, $0.1688\pm 0.0016$ for the
$K^+\pi^-\pi^0$ mode, and $0.2186\pm 0.0024$ for the
$K^+\pi^-\pi^+\pi^-$ mode. The efficiencies are determined using a
detailed GEANT-based Monte Carlo simulation~\cite{ref:geant}, and the
errors quoted are due to Monte Carlo statistics.  

The number of data events that satisfy the selection criteria, $N_e$,
is 1221 in the $K^+\pi^-$ mode, 5249 in the $K^+\pi^-\pi^0$ mode , and
7353 in the $K^+\pi^-\pi^+\pi^-$ mode.  The fraction of signal events
in the data samples is found mode-by-mode using an unbinned maximum
likelihood fit. We define the likelihood function
\begin{equation}
{\cal L} = \prod_{e=1}^{N_e} \left[\sum_{t=1}^7 
	{\cal P}_t(e) f_t\right], \label{eq:likelihood}
\end{equation}
where ${\cal P}_t(e)$ is the normalized probability density function
(PDF) for events of type $t$, evaluated on event $e$, and $f_t$ is the
fraction of such events in the data sample. The seven event types in
the sum are 1)~signal, 2)~$B^+\rightarrow \bar D^0 \pi^+$, 3)~$B
\rightarrow \bar D^*\pi^+ + \bar D^0\rho^+$, 4)~a hard kaon
or 5)~pion in combinatoric $B\bar B$ events with a
misreconstructed $\bar D^0$, and 6)~a hard kaon or 7)~pion in
continuum events. The fit maximizes ${\cal L}$ by varying the seven
fractions, $f_t$, subject to the constraint $\sum_t f_t = 1$.

The PDF's are analytic, six-dimensional functions of the variables
$\Delta E$, $dE/dx$ of the hard kaon candidate, $M(D)$, $M_{bc}$,
${\cal F}$, and $\cos\theta_B$.  The PDF's are mostly products of six
one-dimensional functions, except for correlations between $\Delta E$,
$M(D)$, and $M_{bc}$ in the $B^+\rightarrow \bar D^0 K^+$ and
$B^+\rightarrow \bar D^0 \pi^+$ PDF's.

The $dE/dx$ distributions of $K^\pm,\pi^\pm$ are parameterized
using a Gaussian distribution, whose parameters depend linearly on the
track momentum. The parameterization is determined by studying pure
samples of kaons and pions in data, tagged in the decay chain
$D^{*+}\rightarrow D^0 \pi^+$, $D^0 \rightarrow K^-\pi^+$.
The parameterization in the other variables is obtained from the
off-resonance data for the continuum PDF's and from Monte Carlo for
the $B\bar B$ \/ PDF's.  

The distribution of $B^+\rightarrow \bar D^0K^+$ and $B^+\rightarrow
\bar D^0\pi^+$ events in $\Delta E-M(D)-M_{bc}$ space is parameterized
using the sum of two three-dimensional Gaussians, which are rotated to
account for correlations.  For $B\rightarrow \bar D^*\pi^+ + \bar
D^0\rho^+$ events we use the sum of two Gaussians to parameterize the
$M_{bc}$ and $\Delta E$ distributions, and a Gaussian plus a
bifurcated Gaussian for the $M(D)$ distribution. These distributions
are essentially uncorrelated due to the requirement $\Delta E >
-100$~MeV. For $B\bar B$ events with a misreconstructed $\bar D^0$ we
use a third-order polynomial to parameterize the $\Delta E$
distribution, and a first-order polynomial plus a Gaussian for the
$M(D)$ distribution. The Gaussian is about three times broader than
the $M(D)$ resolution, and models the peaking which arises due to the
selection of the best candidate in the event. The $M_{bc}$
distribution is parameterized using the Argus
function~\cite{ref:argus-function} $f(M_{bc})\propto
M_{bc}\sqrt{1-(M_{bc}/E_b)^2}\exp[-a(1-(M_{bc}/E_b)^2)]$, plus a
Gaussian, which reflects mostly $B\rightarrow \bar D^{(*)}\pi^+$ or
$B^+\rightarrow\bar D^0\rho^+$ events in which we misreconstruct a
$\bar D^0$.

We use a first-order polynomial to parameterize the $\Delta E$
distribution of continuum events, and a first-order polynomial plus a
Gaussian for their $M(D)$ distribution. The Gaussian peaking is due
both to real $\bar D^0$'s and to the selection of the best candidate
in the event.  The $M_{bc}$ distribution is parameterized using an
Argus function whose sharp edge is smeared by adding a bifurcated
Gaussian to account for the beam energy spread.  We use the function
$1-\xi\cos^2\theta_B$ to parameterize the $\cos\theta_B$
distributions, and bifurcated Gaussians for the $\cal F$
distributions.

\begin{figure}[p]
\centering
\leavevmode
\epsfxsize=3.25in
\epsffile{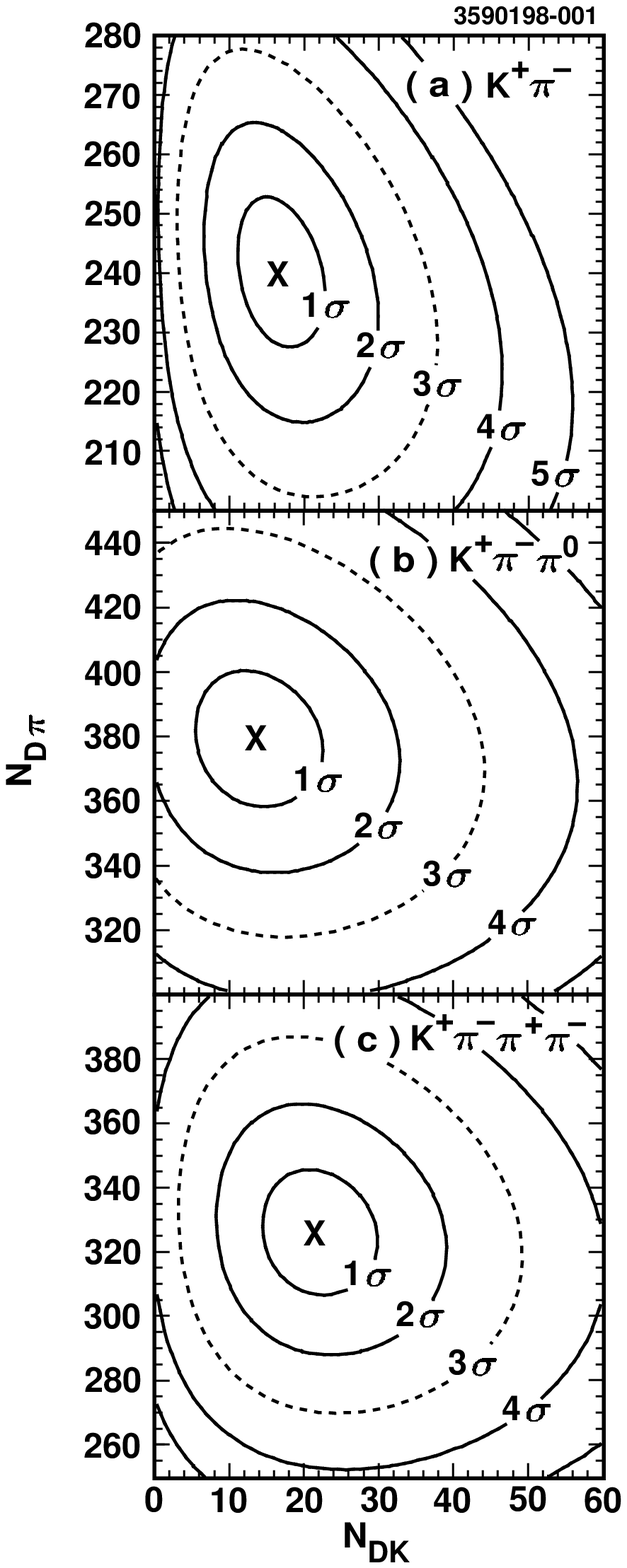}
\caption{Contour plots of $-2\ln {\cal L}$ as a function of $N_{DK}$
and $N_{D\pi}$, the number of $B^+ \rightarrow \bar D^0 K^+$ and $B^+
\rightarrow \bar D^0 \pi^+$ events found in the fit, respectively. The
dashed line marks the $3\sigma$ contour. (a) $\bar D^0\rightarrow
K^+\pi^-$, (b) $\bar D^0\rightarrow K^+\pi^-\pi^0$, (c) $\bar
D^0\rightarrow K^+\pi^-\pi^+\pi^-$. Note that the $N_{D\pi}$ axis has
a suppressed zero.}
\label{fig:contour}
\end{figure}

The results of the maximum likelihood fits are summarized in
Table~\ref{tab:results}. Averaging over the three modes, we find
$R\equiv{\cal B}(B^+\rightarrow \bar D^0 K^+) / {\cal
B}(B^+\rightarrow \bar D^0 \pi^+) = 0.055 \pm
0.014$~(statistical). This is consistent with the value $(f_K /
f_\pi)^2 \tan^2 \theta_c \approx 0.07$, expected from factorization,
with $a_2\ll a_1$~\cite{ref:factorization}. The $\chi^2$ of the
average is $1.2$ for two degrees of freedom, indicating the
consistency among the results obtained with the three decay modes. To
illustrate the significance of the signal yield, contour plots of
$-2\ln {\cal L}$ vs. the number of $B^+\rightarrow \bar D^0 K^+$
and $B^+\rightarrow \bar D^0 \pi^+$ events are shown in
Figure~\ref{fig:contour}. The curves represent $n\sigma$ contours,
corresponding to the increase in $-2\ln {\cal L}$ by $n^2$ over the
minimum value.

\begin{table}[hbp]
\begin{center}
\caption{Results of the maximum likelihood fits. $N_{DK}$ and
	$N_{D\pi}$ are the numbers of $B^+\rightarrow \bar D^0 K^+$
	and $B^+\rightarrow \bar D^0 \pi^+$ events found in the fit,
	respectively. Errors are statistical only.  The statistical
	significance of the signal yield is determined from $-2\ln
	{\cal L}$ by fixing the number of signal events at zero and
	refitting the data.  }
\label{tab:results}
\begin{tabular}{lccc}
Mode:       & $K^+\pi^-$ & $K^+\pi^-\pi^0$ & $K^+\pi^-\pi^+\pi^-$  \\
\hline
$N_{DK}$  & $16.5 \pm 5.9$ & $13.5 \pm 8.7$ & $21.5 \pm 7.8$  \\
$N_{D\pi}$ & $240  \pm 15$  & $379  \pm 22$  & $326  \pm 20$   \\
$N_{DK}$ significance &
	 $4.2\sigma$ & $1.8\sigma$ & $3.8\sigma$ \\
$N_{DK}/N_{D\pi}$ &
	$0.069 \pm 0.026$ & $0.035 \pm 0.023$ & $0.066 \pm 0.025$ \\
\end{tabular}
\end{center}
\end{table}

\begin{figure}
\centering
\leavevmode
\epsfxsize=3.25in
\epsffile{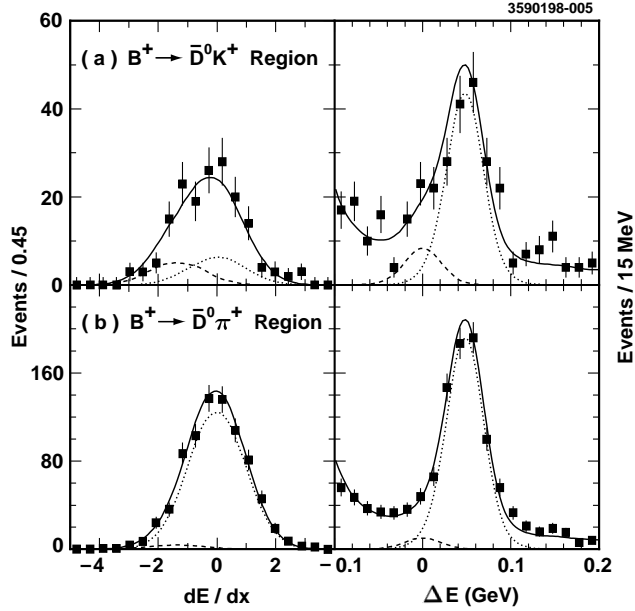}
\caption{Projections onto the $dE/dx$ and $\Delta E$ axes of the
	data (points) and fit function (solid curves), summed over the
	three modes. The dashed and dotted curves are the
	$B^+\rightarrow \bar D^0K^+$ and $B^+\rightarrow \bar
	D^0\pi^+$ contributions to the fit functions,
	respectively. (a) $B^+\rightarrow \bar D^0 K^+$ region. (b)
	$B^+\rightarrow \bar D^0 \pi^+$ region.}
\label{fig:projections}
\end{figure}

The quality of the fit is illustrated in
Figure~\ref{fig:projections}a, showing projections of the data onto
$dE/dx$ and $\Delta E$ for events in the $B^+\rightarrow \bar D^0
K^+$ region, defined by ${\cal F} < 1.6$, $|M_{bc} - 5280 \ {\rm MeV}|
< 5\ {\rm MeV}$, $|M(D) - 1864.5 \ {\rm MeV}| < 20 \ {\rm MeV}$, $-50
< \Delta E < 10 \ {\rm MeV}$, $dE/dx < -0.75$. Requiring that
events fall within this $B^+\rightarrow \bar D^0 K^+$ region reduces
the signal efficiency by about 50\%, but strongly suppresses the
background. Overlaid on the data are projections of the fit
function. The fit function is the sum of the PDF's, each weighted by
the number of corresponding events found in the fit and multiplied by
the efficiency of the corresponding event type to be in the
$B^+\rightarrow \bar D^0 K^+$ region.  In
Figure~\ref{fig:projections}b we show projection plots for events in
the $B^+\rightarrow \bar D^0 \pi^+$ region, defined by $0 < \Delta E <
100 \ {\rm MeV}$, $|dE/dx| < 2.5$, and with the same requirements
on $\cal F$, $M_{bc}$ and $M(D)$ as in the $B^+\rightarrow \bar D^0
K^+$ region. These projections demonstrate that the fit function
agrees well with the data in the regions most highly populated by
signal and the most pernicious background, and provides confidence in
our modelling of the tails of the $B^+\rightarrow \bar D^0 \pi^+$
distributions.

\begin{figure}
\centering
\leavevmode
\epsfxsize=3.25in
\epsffile{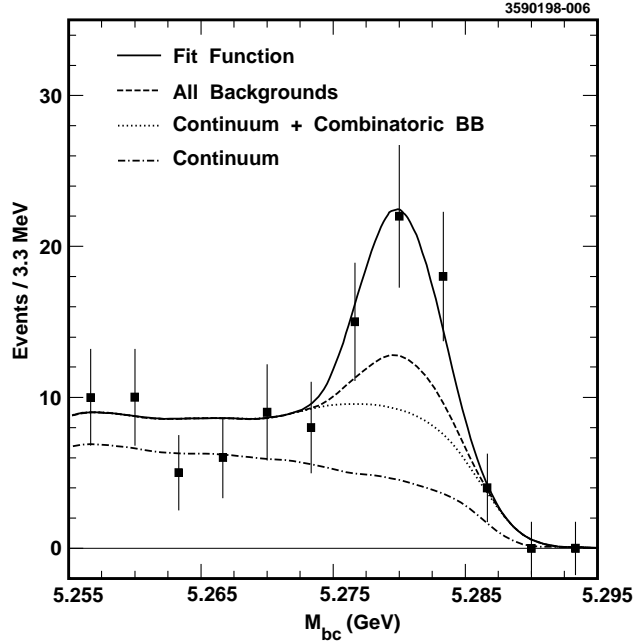}
\caption{Projections onto the $M_{bc}$ axis of the data (points) and
	fit function (solid curve) in the $B^+\rightarrow \bar
	D^0 K^+$ region, summed over the three modes. Also shown are
	separate background contribtions to the fit function:
	Continuum, continuum plus combinatoric $B\bar B$ events
	with a misreconstructed $\bar D^0$, and all backgrounds
	(including $B^+\rightarrow \bar D^0 \pi^+$).}
\label{fig:projections-mb}
\end{figure}

Projections onto $M_{bc}$ for events in the signal region
(Figure~\ref{fig:projections-mb}) illustrate the relative
contributions and distributions of signal and background events. Only
$B^+\rightarrow \bar D^0K^+$ and $B^+\rightarrow \bar D^0\pi^+$ events
peak significantly around $M_{bc} = M_B$, despite the selection of the
best candidate in the event.


We conduct several tests to verify the consistency of our result.  The
fit is run on off-resonance data and on Monte Carlo samples containing
the expected distribution of background events with no signal. In both
cases the signal yield is consistent with zero. We also fit the data
without making use of $\cal F$ or $dE/dx$, and obtain results
consistent with those of Table~\ref{tab:results}, with increased
errors. We find the branching fraction ${\cal B}(B^+\rightarrow
\bar D^0 \pi^+) = (4.82 \pm 0.19 \pm 0.31)\times 10^{-3}$, in
agreement with previous CLEO measurements~\cite{ref:b2dpi}. The ratio
between the $B \rightarrow \bar D^*\pi^+ + \bar D^0\rho^+$
and $B^+\rightarrow \bar D^0 \pi^+$ yields obtained from the fit
is consistent with the measured branching fractions of these
decays~\cite{ref:pdg}. In addition, our $B^+\rightarrow \bar
D^0K^+$ result is consistent with that of a simpler, though less
sensitive method, used to analyze the same data~\cite{ref:warsaw}.


Many systematic errors cancel in the ratio $R$. We assess systematic
errors due to our limited knowledge of the PDF's by varying all the
PDF parameters by $\pm1$~standard deviation in the basis in which they
are uncorrelated, where the magnitude of a standard deviation is
determined by the statistics in the data or Monte Carlo sample used to
evaluate the PDF parameters. The systematic error in $R$ due to Monte
Carlo statistics is 0.0033. The error due to statistics in the data
sample used to parameterize the $dE/dx$ distributions is 0.0028,
and the error due to statistics in the off-resonance data sample is
0.0017. 
We assign a systematic error of 0.0005 due to the uncertainty
in the average beam energy, which we estimate to be $\pm0.16$~MeV by
using the peak of the $M_{bc}$ distribution of $B^+\rightarrow \bar
D^0 \pi^+$ events. The total systematic error is 0.0047.


In summary, we have observed the decay $B^+\rightarrow \bar D^0
K^+$ and determined the ratio of branching fractions 
\begin{equation}
R = {{\cal B}(B^+\rightarrow \bar D^0 K^+) \over 
	{\cal B}(B^+\rightarrow\bar D^0 \pi^+)}
	 = 0.055 \pm 0.014 \pm 0.005. 
\end{equation}
Combining this
result with the CLEO~II measurement~\cite{ref:b2dpi} ${\cal
B}(B^+\rightarrow \bar D^0 \pi^+) = (4.67 \pm 0.22 \pm
0.40)\times 10^{-3}$, we obtain ${\cal B}(B^+\rightarrow \bar D^0
K^+) = (0.257 \pm 0.065 \pm 0.032)\times 10 ^{-3}$.

We gratefully acknowledge the effort of the CESR staff in providing us with
excellent luminosity and running conditions.
This work was supported by 
the National Science Foundation,
the U.S. Department of Energy,
Research Corporation,
the Natural Sciences and Engineering Research Council of Canada, 
the A.P. Sloan Foundation, 
and the Swiss National Science Foundation.

\end{document}